\documentclass[10pt]{iopart}

\usepackage{epsf}

\begin{document}

\title{Equilibrium and Aging Dynamics of Simple Models for Glasses}

\author{A. Crisanti\dag\ and F. Ritort\ddag}

\address{\dag\ Dipartimento di Fisica, Universit\`a di Roma ``La Sapienza''
         and \\
         Istituto Nazionale Fisica della Materia, Unit\`a di Roma\\
         P.le Aldo Moro 2, I-00185 Roma, Italy}

\address{\dag\ Physics Department, Faculty of Physics \\
         University of Barcelona, Diagonal 647, 08028 Barcelona,  Spain
        }

\begin{abstract}
We  analyze the properties of the energy landscape of {\it finite-size} 
fully connected $p$-spin-like models. 
In the thermodynamic limit the high temperature phase is described by 
the schematic Mode Coupling Theory of super-cooled liquids.
In this limit the barriers between 
different basins are infinite below the critical dynamical temperature
the ergodicity is broken on in infinite times. 
We show that {\it finite-size} fully connected $p$-spin-like models,
where activated processes are possible, 
do exhibit properties typical of real
super-cooled liquid when both are near the critical glass transition.
Our results support the conclusion that fully-connected  
$p$-spin-like models are the natural statistical mechanical models for 
studying the glass transition in super-cooled liquids. 
\end{abstract}


\maketitle

\section*{Introduction}
In recent years a significant effort has been devoted to understand
slow relaxation dynamics observed in many, apparently unrelated,
systems such as structural glasses, spin glasses, disordered 
and granular materials or proteins among others. 
In such systems the characteristic relaxation time may change of
many orders of magnitude if the external parameters, e.g. the temperature
$T$, are slightly varied. As a consequence correlations display 
non-exponential behavior, and equilibration processes slow down 
giving rise to non-equilibrium phenomena known as aging. 

The common denominator which makes all these systems displaying
similar behavior near the (dynamical) critical temperature is the
complexity of the energy landscape.  The trajectory of the
representative point in the configuration space can be viewed as a
path in a multidimensional potential energy surface \cite{GO}. The
dynamics is therefore strongly influenced by the topography of the
potential energy landscape: local minima, barriers, basins of
attraction an other topological properties all influence the dynamics.

The potential energy surface of a super-cooled liquid 
contains a large number of local minima, called {\it inherent structures}
(IS) by Stillinger \cite{S95}. 
All states that under local energy minimization will flow into the same IS
define the {\it basin} of the IS (valley). With this pictures in mind
the time evolution of the system can be seen as the result of two
different processes: thermal relaxation into basins
({\it intra-basin} motion)  and 
thermally activated potential energy barrier crossing between different
basins ({\it inter-basin} motion). 
When the temperature is lowered down to the order of the critical 
Mode Coupling Theory (MCT) temperature $T_{MCT}$ the inter-basin 
motion slows down and the relaxation dynamics is dominated by 
the slow 
thermally activated crossing of potential energy barriers \cite{SSDG99}.
If the temperature is further reduced the
relaxation time eventually becomes of the same order of the observation 
time and the system falls out of equilibrium since there is not
enough time to cross barriers and equilibrate. 
This define the ``experimental''glass transition temperature $T_g$.

The regime between $T_{MCT}$ and $T_g$ cannot be described by the MCT 
since it neglects activated processes responsible for barrier crossing. 
In MCT the relaxation time diverges at $T_{MCT}$, leading to $T_g=T_{MCT}$, 
and the dynamics
remains confined into a single basin forever. Attempts to overcome these 
difficulties in MCT have been done, but probably the most clear picture 
comes from some spin-glass models. 
The essential features of MCT for glass-forming systems 
are also common to some fully connected spin glass models \cite{KT87},
the most well known being the 
spherical $p$-spin spin glass model \cite{CS92,CHS93}. 
We shall call these models 
{\it mean-field $p$-spin-like} glass models. 
The central point is that
near $T_{MCT}$ the behavior of the system is mainly due to the IS 
organization (density, basins, barriers and so on), so that all systems
with similar IS structure should have similar critical behavior.

In the thermodynamic limit the high-temperature phase --
paramagnetic in the spin-glass language and liquid in glass language --
of mean-field $p$-spin-like
is described by the schematic MCT for super-cooled liquids 
\cite{G84,CHS93}. 
As a consequence at the critical temperature, called $T_D$ in $p$-spin
language, an ergodic to non-ergodic transition 
takes place. 
Below this temperature the system is dynamically confined to a metastable
state (a basin) \cite{CS95} since
relaxation to true equilibrium can only take place via 
activated processes, absent in mean-field models.
Therefore in mean-field models, similar to what happens in MCT, 
at $T_D$ the relaxation time diverges. 
For these systems, nevertheless, it is known that the true equilibrium 
transition to a low temperature phase occurs below $T_D$ at the static 
critical temperature $T_c$, also denoted by $T_{1rsb}$ \cite{CS92}. 
This is the analogous of the Kauzmann temperature $T_k$ for liquids.
The glass transition temperature $T_g$ of real systems sits somewhere
in between $T_c$ and $T_D$. 
This transition, obviously, cannot be reached even on infinite time in 
mean-field models.

Despite these difficulties mean-field models, having the clear advantage
of being analytically tractable, have been largely 
used to study the properties of fragile glassy systems, 
especially between the dynamical
temperature $T_D$ and the static temperature $T_c$ 
where a real thermodynamic phase transition
driven by the collapse of the configurational entropy takes place.  
The picture that emerges is however not complete since
activated process cannot be captured by mean-field models. 
Therefore the relevance of mean-field results for real systems
it is still not completely stated. 
Let us also
remark that, despite the large amount of analytical work devoted to
the study of the static as well as dynamical properties in the
$N\to\infty$ limit, much less is known concerning the finite $N$
behavior.

In this work we investigate numerically {\it finite-size} 
fully-connected $p$-spin-like models, 
where activated processes {\it are} present.  
Comparing our results with the observed behavior of super-cooled liquids 
near $T_{MCT}$ we conclude that, once activated process are allowed, 
mean-field $p$-spin-like models are highly valuable for a deep 
understanding of the glass transition in real systems.

All results reported here are for the Ising-spin Random
Orthogonal Model (ROM) \cite{MPR94,PP95}, 
however similar results are obtained using other $p$-spin-like models, 
as for example the Bernasconi Model and the Ising $p$-spin model. 
The advantage of ROM lies in its interaction term which is two-body, 
at difference  with the $p$-body interaction of $p$-spin models, 
reducing computer memory problems.
Moreover the use of Ising spins instead of continuous spins, as for 
the spherical $p$-spin model, allows for a larger configuration entropy and
faster algorithms. Preliminary results for the spherical $p$-spin model
are in qualitative agreement with those reported here.

\section{Thermodynamics of Inherent Structures: How to evaluate the configurational entropy}

The Random Orthogonal Model model is defined by the Hamiltonian
\cite{MPR94,PP95},
\begin{equation}
\label{eq:ham}
  H = - 2 \sum_{ij} J_{ij}\, \sigma_i\, \sigma_j 
\end{equation}
where $\sigma_i=\pm 1$ are $N$ Ising spin variables, 
and $J_{ij}$ is a $N\times N$ random 
symmetric orthogonal matrix with $J_{ii}=0$.  Numerical simulations are
performed using the Monte Carlo method with the Glauber algorithm.
For $N\to\infty$ this model has the
same thermodynamic properties of the $p$-spin model:
a dynamical transition at $T_D=0.536$, 
with threshold energy per spin $e_{th} = E_{th}/N = -1.87$, 
and a static transition at
$T_c=0.25$, with critical energy per spin $e_{1rsb}= -1.936$ 
\cite{MPR94,PP95}. 

The TAP analysis \cite{CS95,PP95} reveals 
that the phase space visited
is composed by an exponentially large (in $N$) number 
of different basins, each labelled
by the value of the energy density $e$ at $T=0$,
separated by infinitely large (for $N\to\infty$) barriers. 
The free energy of the $e$-TAP solution describes the 
thermodynamics within the basins labelled by 
$e$ and at $T=0$ coincides with the local minima potential energy, i.e.,  
the  IS of the system. 
The dynamical transition is associated with IS having the 
largest basin of attraction for $N\to\infty$, while the static transition
with IS having the lowest accessible free energy \cite{CS95,KW87}. 

In the mean-field limit, the allowed values of $e$ are between 
$e_{1rsb}$ and $e_{th}$. 
Solutions with $e$ larger than $e_{th}$ are unstable (saddles),
while solutions with $e$ smaller than $e_{1rsb}$ have negligible statistical
weight. Moreover in the $N\to\infty$
limit IS with $e=e_{th}$ attract most 
(exponentially in $N$) of the states and dominate the behavior of the 
system. Other IS are irrelevant for $N\to\infty$.

For finite $N$ the scenario is different since not only the basins of 
IS with $e < e_{th}$ acquire statistical weight,
but it may happen that solutions with $e>e_{th}$ 
and few negative directions (saddles with few downhill directions) 
become stable, simply because there are not 
enough degrees of freedom to hit them. 

\begin{figure}[hbt]
\epsfxsize=10cm\epsfysize=7cm
\epsfbox{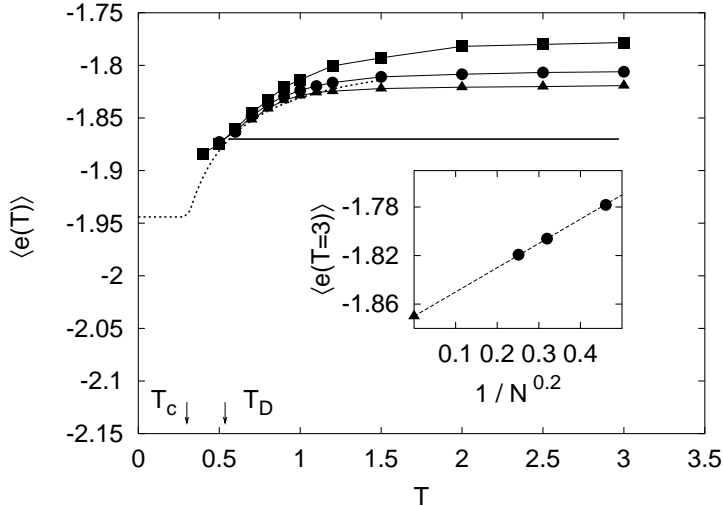}
\caption{Temperature dependence of $\langle e(T)\rangle$ for
         $N=48$ (square), $N=300$ (circle) and $N=1000$ (triangle). 
	The average is over $10^3$ different equilibrium
	configurations at temperature $T$. 
	 The horizontal line is the $N\to\infty$ limit.
         The arrows indicate the critical temperatures $T_D$  
         and $T_c$ (see text). 
         The dotted line is the curve obtained from the
         configurational entropy for large $N$.
         Inset: size dependence of 
         $\langle e(T=3)\rangle$ as a function $N^{-\alpha}$ 
         with $\alpha=0.2$ extrapolated down to the $N\to\infty$ 
         theoretical result $-1.87$ (triangle).
}
\label{fig:e-is}
\end{figure}

To get more insight the IS-structure of finite systems we follow 
Stillinger and Weber \cite{SW82} and decompose the
partition sum into a sum over basins of different IS and
a sum within each basin. Collecting all IS with the same energy
$e$, denoting with $\exp [N s_c(E)]\,de$ the number of IS 
with energy between 
$e$ and $e+de$, and shifting the energy of each basin with
that of the associated IS,
the partition sum can be rewritten as \cite{SW82}
\begin{equation}
\label{eq:part}
  Z_N(T)\simeq \int d e \exp\, N\,\left[-\beta e + s_c(e)
                                     -\beta f(\beta,e)
                               \right]
\end{equation}
where $f(\beta,e)$ can be seen as the free energy density of the
system when confined in one of the basin associated with IS
of energy $e$. The function $s_c(e)$ is the {\it configurational
entropy density} also called {\it complexity}. From eq. (\ref{eq:part}) 
we easily obtain the probability that an equilibrium
configuration at temperature $T=1/\beta$ lies in a basin associated with
IS of energy between $e$ and $e+de$:
\begin{equation}
\label{eq:prob}
 P_N(e,T) =  \exp\, N\,\left[-\beta e + s_c(e)
                                     -\beta f(\beta,e)
                               \right] / Z_N(T).
\end{equation}
Taking the $N\to\infty$ limit we recover the mean-field
results \cite{CS95,MPR94,PP95}. 

>From the partition function (\ref{eq:part}) we can easily compute the average
internal energy density given by
$u(T) = \langle e +\partial (\beta f) / \partial \beta \rangle =$
$\langle e(T) \rangle + \langle \Delta e(T)\rangle$. 
The first term is the average energy of the IS relevant for the thermodynamics
at temperature $T$, while  the second term is the contribution 
from fluctuations inside the basin of the IS.
The average is taken with the weight (\ref{eq:prob}). 

\begin{figure}[hbt]
\epsfxsize=10cm\epsfysize=8cm
\epsfbox{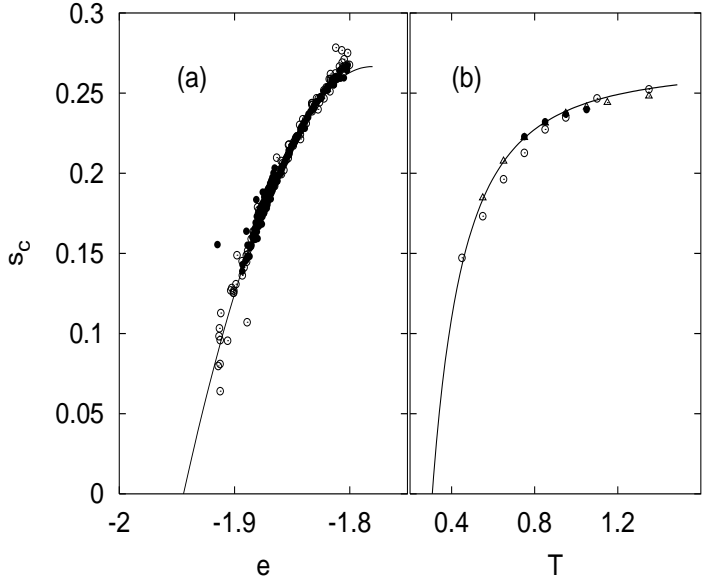}
\caption{(a) Configurational entropy as a function of energy.
         The data are from system sizes $N=48$ (empty circle) and $N=300$
         (filled circle), and temperatures $T=0.4$, $0.5$, $0.6$, $0.7$, 
         $0.8$, $0.9$ and $1.0$. For each curve the unknown constant has
         been fixed to maximize the overlap between the data and
         the theoretical result \protect\cite{PP95}. The line is the
	 quadratic best-fit.
         (b) Configurational entropy density as a function of temperature.
         The line is the result from the best-fit of $s_c(e)$ 
         while the symbols are the results from the temperature integration
         of eq. (\protect\ref{eq:entro}) for $N=48$ (empty circle), 
         $N=300$ (empty triangle) and $N=1000$ (filled circle). 
}
\label{fig:compl}
\end{figure}

Since we are interested into the IS structure we shall concentrate on
$\langle e(T) \rangle$.
In the limit $N\to\infty$ the only relevant IS are those with $e=e_{th}$,
and $\lim_{N\to\infty} \langle e(T) \rangle = e_{th}$ for any $T>T_D$. 
To measure $\langle e(T) \rangle$ for finite $N$ we perform the following 
experiment. First we equilibrate the system at a given temperature $T$, 
then starting from an equilibrium configuration we
instantaneously quench it down to $T=0$. This is obtained by decreasing the
energy along the steepest descent path. In this way we can identify the
energy of the IS visited by the equilibrium trajectory. 
The experiment is repeated several times starting from
uncorrelated equilibrium configurations at $T$ and the average IS energy
is computed. 
In figure \ref{fig:e-is} we report $\langle e(T)\rangle$ as a function 
of temperature $T$ for system sizes $N=48$, $300$ and $1000$. 
As expected, as $N$ increases $\langle e(T)\rangle$
tends towards $e_{th}$. From the numerical data we found that the plateau energy,
approaches $e_{th}$ with the power law
$\langle e_{plateau}\rangle - e_{th} \sim N^{-0.2}$, see inset 
Fig. \ref{fig:e-is}.
Note that 
since $N$ is finite we can equilibrate the system also 
below $T_D$, down to the glassy transition $T_g(N)$, about $0.35$ for $N=48$ 
and $0.5$ for $N=300$, 
below which the system falls out of equilibrium \cite{Note1}.

The figure shows that for finite $N$ and $T$ not too close to
$T_D$ the thermodynamics is dominated by IS with $e>e_{th}$. This is more 
evident from the (equilibrium) probability distribution of $e$
since it is centered about $\langle e(T)\rangle$ indicating that IS
with $e\simeq \langle e(T)\rangle$ have the largest basins.
This scenario has been also observed in real 
glass-forming systems\cite{OTW88,SDS98,SST98,Parisi,KST99}.

>From the knowledge of IS-energy distribution we can reconstruct the
complexity $s_c(e)$. In 
the temperature range where eq. (\ref{eq:prob}) is valid, we have
\begin{equation}
\label{eq:compl}
  s_c(e) = \ln P_N(e,T) + \beta e + \beta f(\beta,e) +
               \ln Z_N(T)
\end{equation}
If energy dependence of $f(\beta,e)$ can be neglected, then it is possible 
to superimpose the curves for different temperatures. 
The resulting curve is, except for
an unknown constant, the complexity $s_c(e)$. 
The curves obtained for system sizes $N=48$ and $300$ and various temperatures
between  $T=0.4$ and $T=1.0$ are shown 
in figure \ref{fig:compl} (a). The data collapse is
rather good for $e < -1.8$. Above the curves cannot be superimposed anymore
indicating that the energy dependence of $f(\beta,e)$ cannot be neglected.
In liquid this is called the anharmonic threshold
\cite{SKT99,BH99}.
To compare the result with the known analytical predictions for the ROM
each curve in the figure has been translated to maximize the overlap with the
theoretical prediction for $s_c(e)$ \cite{PP95}.  
The dotted line is the quadratic best 
fit from which we can estimate the critical energy $e_c\simeq -1.944$
as the value where $s_c(e)$ vanishes 
in good agreement with the theoretical result $e_{1rsb}=-1.936$ \cite{PP95}.

Direct consequence of $f(\beta,e) \simeq f(\beta)$ for $e<-1.8$ is that
in this range the partition function can be written as the product of an
intra-basin \cite{Note2} contribution [$\exp(-N\beta f$)] 
and of a configurational contribution which
depends only on the IS energy densities distribution.
The system can then be considered as composed by two independent subsystems:
the intra-basin subsystem describing the equilibrium when confined within 
basins, and the IS subsystem describing equilibrium via activated processes
between different basins.
As the temperature is lowered
and/or $N$ increased the two processes get more separated in time and the 
separation into two subsystems becomes more and more accurate. 
A scenario typical of super-cooled liquids near the MCT 
transition \cite{ST97,SSDG99}.

The form of $f(\beta)$ for the specific system can be computed 
studying the motion near the IS, for example
using an harmonic approximation \cite{SKT99,BH99}. However, this usually 
gives only a small corrections to thermodynamic quantities for $T$ close to 
$T_D$ \cite{SKT99,BH99} and we do not consider it here.

Another important consequence of the separation into two subsystems 
is that in eq. (\ref{eq:prob}) $f(\beta,e)$ can be neglected since 
it cancels with the equal term coming from the denominator.
Therefore from the knowledge of $s_c(e)$ we can easily compute 
the average IS energy density $\langle e(T)\rangle$. 
For large $N$ this is given by the saddle point estimate, 
see eq. (\ref{eq:prob}):
\begin{equation}
\label{eq:eav}
 \max_{e}\, \left[-\beta\, e + s_c(e) \right].
\end{equation}
The result obtained using for $s_c(e)$ 
the quadratic best-fit of figure \ref{fig:compl} (a)
is the dotted line shown in Fig. \ref{fig:e-is}. The agreement with the 
direct numerical data is good already for $N=300$.
>From the form of $\langle e(T)\rangle$ for large $N$ 
we can identify the static critical temperature $T_c\simeq 0.3$
as the temperature below which $\langle e(T)\rangle$ remains constant,
not far from the theoretical results $T_c=0.25$ \cite{PP95}. 

To have a check of our results we have computed the configurational 
entropy density using a different approach \cite{SKT99}. 
When the IS subsystem is in 
thermal equilibrium then the temperature dependence of the configurational
entropy density can be evaluated from the thermodynamic relation
\begin{equation}
\label{eq:entro}
 \frac{d\, s_c(T)}{d\, \langle e(T)\rangle} = \frac{1}{T}
\end{equation}
integrating the $T$ dependence of $d\,\langle e(T)\rangle / T$.
Using the data of figure \ref{fig:e-is} we obtain the curves shown in figure
\ref{fig:compl} (b). The line is the result valid for large $N$ obtained from 
the quadratic best fit of $s_c(e)$ [figure \ref{fig:compl} (a)].
The agreement for $N=300$ and $1000$ is rather good. 
Note that increasing $N$ reduces the IS energy range
explored by the system for a given fixed Montecarlo simulation length. 
This in turn reduces the temperature range in 
figure \ref{fig:compl} (b). 

To have a statistical description of the energy landscape we also
investigated cross correlations among IS and equilibrium configurations
at $T$.  It emerges that while the equilibrium configuration is highly
correlated with the corresponding IS, different IS are
uncorrelated. Moreover the probability distribution of IS-IS overlaps is
a Gaussian centered in zero and variance which goes to zero as $N$
increases. The analysis of triplet of IS does not reveal any particular
organization of states. This result, characteristic of the Random Energy
Model \cite{DERRIDA}, is known to hold also for multispin interaction
spin-glass models.

\begin{figure}[hbt]
\epsfxsize=10cm\epsfysize=8cm
\epsfbox{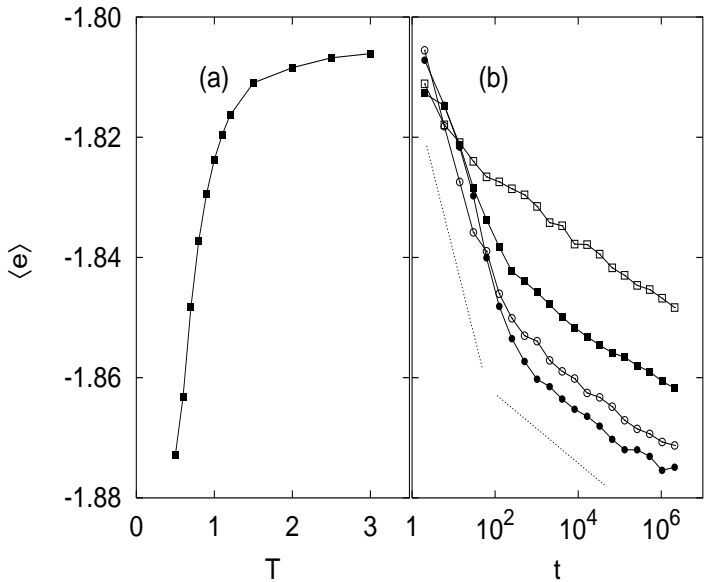}
\caption{Average inherent structure energy in equilibrium as a function
         of temperature $(a)$ and as a function of time during 
         the non-equilibrium process $(b)$. The system size is $N=300$,  
         $T_i= 3.0$ and (top to bottom)
         $T_f= 0.1$, $0.2$, $0.3$ and $0.4$ (panel b). 
}
\label{fig:e-aging}
\end{figure}

\section{Non-equilibrium behavior: the role of activated processes}

More informations on the IS structure can be obtained from
non-equilibrium relaxation processes.  To study the non-equilibrium
dynamics we quench at time zero the system from an initial equilibrium
configuration at temperature $T_i>T_g$ to a final temperature
$T_f<T_g$ and study the evolution of the average IS energy per spin
$\langle e(t)\rangle$ as function of time.  This is obtained by
regularly quenching the system down to $T=0$ to calculate the
instantaneous IS energy.  The result for a system of $N=300$ spins and
initial temperature $T_i= 3$ is shown in figure \ref{fig:e-aging} (b)
for final temperature $T_f=0.1$, $0.2$, $0.3$ and $0.4$. The average
is over different equilibrium initial configurations at $T_i$.

The analysis of the figure reveals that the relaxation process can be 
divided into two different regimes. A first regime
independent of $T_f$, and a second regime independent of 
both $T_i$ and $T_f$. The final temperature $T_f$ controls the cross-over 
between the two regimes. A similar behavior has been observed in 
molecular dynamics simulations of super-cooled liquids \cite{KST99}. 
We note that in our case since we use discrete variables, and hence a 
faster dynamics, we do not
have the very-early regime observed in \cite{KST99} where 
$\langle e(t)\rangle$ is almost independent of $t$. 

\begin{figure}[hbt]
\epsfxsize=10cm\epsfysize=8cm
\epsfbox{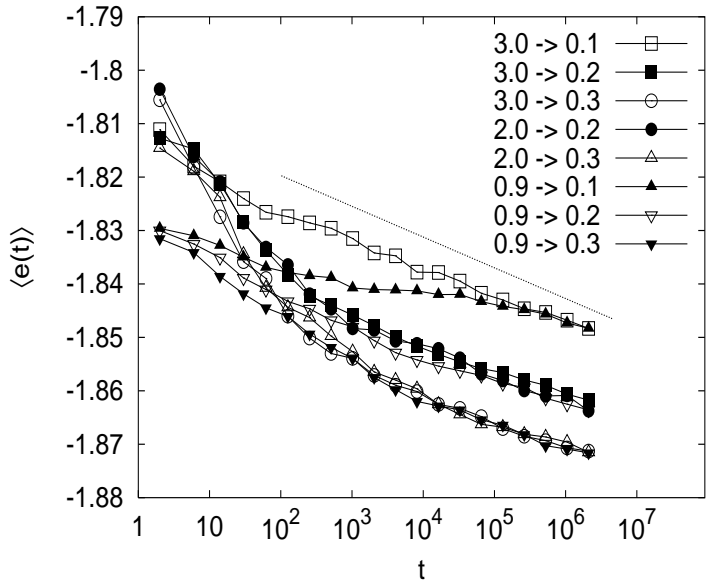}
\caption{Average inherent structure energy as function of time for
         initial temperatures $T_i= 3.0$, $2.0$ and $0.9$, final
         temperatures $T_f= 0.1$, $0.2$ and $0.3$. The average is over
         $300$ initial configurations. The system size is $N=300$. 
         The line denotes the slope $-0.0025$.
}
\label{fig:e-agi2}
\end{figure}

The two regimes are associated with different relaxation processes.
In the first part the system has enough energy and
relaxation is mainly due to {\it path search} out of basins through
saddles of energy lower than $k_{B}T_f$.
This part depends only on the initial 
equilibrium temperature $T_i$ since it sets the initial phase space region.
Different $T_i$ leads to different power law. 
In particular relaxation must slow down as 
$T_i$ decreases since we expect that lower states are surrounded by higher 
barriers.
This expectation is supported by our numerical data. In figure \ref{fig:e-agi2}
we report the behavior of $\langle e(t)\rangle$ as function of time
for different initial temperatures. 
The slowing own of the first 
regime is clearly seen.

During this process the system explores deeper and deeper valleys (basins)
while decreasing its energy. The process stops when all barrier heights
become of $O(k_{B}T_f)$.
>From now on the relaxation proceeds only via activated process. 
A first consequence is that lower the final temperature $T_f$ 
shorter the first relaxation, in agreement with our findings [See figures
\ref{fig:e-aging} and \ref{fig:e-agi2}].

The analysis of 
the distance between the instantaneous system state 
and the corresponding IS, counting  the number of single spin flip 
needed to reach the IS, reveals that 
for all times the systems stays in configurations 
few spin flips away from an IS. The number ranging from $8-9$ 
for short times to $1-2$ at larger times. A similar study starting from 
equilibrium configurations at temperature 
$T_e(\langle e(t)\rangle)$ evaluated comparing panels (a)
and (b) of figure \ref{fig:e-aging} \cite{KST99} leads to similar numbers.
We then conclude that during relaxation the aging system explores 
the same type of minima (and basins) visited in equilibrium
at temperature $T_e$. 
Direct consequence is that 
once the system has reached the activated regime  there cannot be
memory of the initial $T_i$, and all curves with different $T_i$ but
same $T_f$ should collapse for large time [figure \ref{fig:e-agi2}].

\begin{figure}[hbt]
\epsfxsize=10cm\epsfysize=8cm
\epsfbox{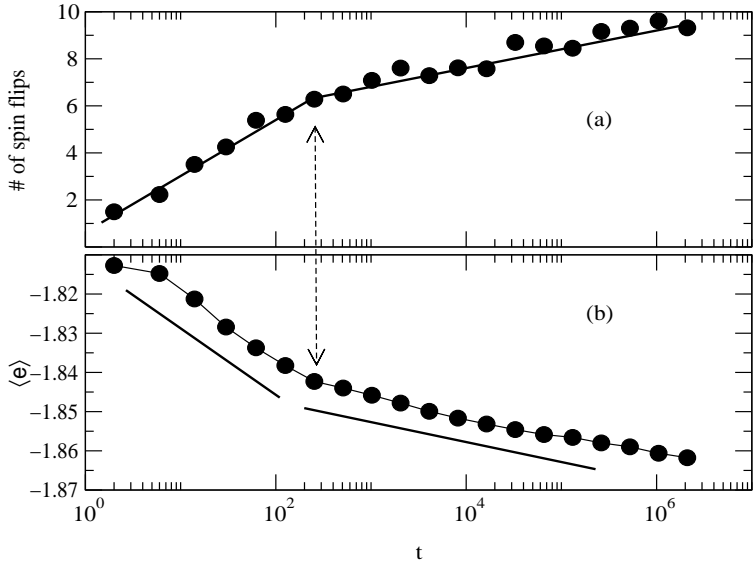}
\caption{$(a)$ Number of single spin flip to reach the nearest saddle from the IS
          as a function of $t$t. 
         $(b)$ $\langle e(t)\rangle$  as a function of $t$during 
         the non-equilibrium process. The system size is $N=300$,  
         $T_i= 3.0$ and $T_f= 0.2$.
        }
\label{fig:flip}
\end{figure}

To have more confidence with this picture of relaxation we have
studied the distance from the instantaneous configurations and the 
nearest saddle to a different basin. This is done by counting the number
of spin flips needed to reach the saddle through the {\it less steep} 
path, i.e., the path that gives the minimum increase of energy at 
each spin flip. 
A typical result is shown in figure \ref{fig:flip}. Note the strong 
slowing down at the ``kink'' where the relaxation law changes.

\begin{figure}[hbt]
\epsfxsize=10cm\epsfysize=8cm
\epsfbox{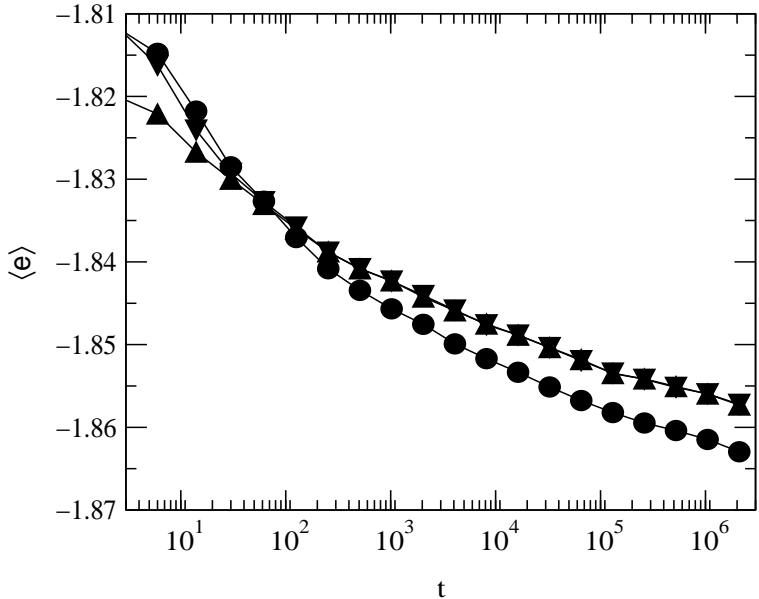}
\caption{ IS energy as function of $t$ for $N=300$,
         $T_i= 3.0$ and $T_f= 0.2$.
         The average is over $300$ initial configurations. 
         Symbols are: circle: IS energy from instantaneous configuration;
                      triangle down: IS energy crossing nearest saddle to
                      instantaneous configuration;
                      triangle up: IS energy crossing nearest saddle to
                      instantaneous IS.
        }
\label{fig:e-flip}
\end{figure}

Finally in figure \ref{fig:e-flip} it is shown the average IS energy of
IS reached by crossing the saddle nearest to the instantaneous configuration.
Note that for short time the crossing leads to IS with similar energies.
By comparison we also report the
analogous curve obtained starting from the instantaneous IS. In 
this case the energy is much lower. The picture changes near the 
kink where both curves merges. 
Moreover for longer times both curves are above the true relaxation 
curve, indicating that the system does not relaxes passing through
the {\it less steep} path.
Indeed 
this is a very special path which may be difficult to find. Higher saddles
require higher activation energy but can be found more easily and 
dominate the relaxation dynamics. 
More detailed studies of barriers heights and exit time are in progress.

\section{Conclusions}

To summarize, in this work we have shown that {\it finite-size}
mean-field $p$-spin-like models are good candidates for studying the glass
transition. The key point is that near the glass
transition the thermodynamics of the systems is dominated by the IS
distributions, therefore all systems with similar IS distributions
should have similar behavior. 
Finite-size mean-field $p$-spin-like models have the double advantage 
of being analytically tractable for $N\to\infty$ and easily simulated 
numerically for finite $N$, offering good models to analyze the glass 
transition. Finally we note that the
analysis presented in this paper opens
the way for the study of generic glass and spin-glass models (such as
the Edwards-Anderson model) where a careful study of the thermodynamics
associated to the IS has never been considered.

\ack
We thank for useful discussions 
C. Donati, U. Marini Bettolo, F. Sciortino and P. Tartaglia.


\end{document}